\documentclass[onecolumn,preprintnumbers,amsmath,aps,
]{revtex4-1}
\usepackage{graphicx}
\usepackage{dcolumn}
\usepackage{bigints}
\usepackage{bm}
\usepackage{subfigure}
\usepackage{color}
\usepackage{microtype}
\usepackage{amsmath}
\usepackage{mathtools}
\usepackage{amssymb}
\begin{document}
\def\eq#1{(\ref{#1})}
\def\fig#1{figure\hspace{1mm}\ref{#1}}
\def\tab#1{table\hspace{1mm}\ref{#1}}
\title{Mimetic-$f(Q)$ gravity: cosmic reconstruction and energy conditions}

\author{Adam Z. Kaczmarek}\email{a.kaczmarek@doktorant.ujd.edu.pl}\email{adamzenonkaczmarek@gmail.com}
\affiliation{Institute of Physics, Faculty of Science and Technology, Jan D{\l}ugosz University in Cz{\c{e}}stochowa, 13/15 Armii Krajowej Ave., 42200 Cz{\c{e}}stochowa, Poland}

\date{\today} 
\begin{abstract}

In this study, we present a novel approach to mimetic gravity incorporating a non-zero nonmetricity tensor with vanishing torsion and curvature, establishing a generalized mimetic-$f(Q)$ gravity framework. Using the Lagrange multiplier method, we have obtained and discussed characteristics of the theory's field equations. In order to study cosmic evolution given by the hybrid scale factor, we implemented the reconstruction method in two different ways. In the first case, we have obtained corresponding Lagrange multiplier $\eta$ and potential $U$ for the specific $f(Q)=f(Q)=Q-6\lambda M^2\big(\frac{Q}{6M^2}\big)^\alpha$ function, while in the second scenario we have recovered $f(Q)$ functional and mimetic potential for the given Lagrange multiplier $\eta_0+\gamma H^2$. Subsequently, we explore the fundamental properties of the $f(Q)=Q-6\lambda M^2\big(\frac{Q}{6M^2}\big)^\alpha$ model and analyse the energy conditions to establish its validity. Our findings indicate that the framework introduced herein allows for the derivation of a wide range of viable cosmological models that satisfy energy constraints that are necessary in description of the accelerated expansion.
\end{abstract}

\maketitle
\section{Introduction}


Recent experiments suggested that almost $95 \%$ of the Universe is consisted from the dark sector, indicating lack of completeness of the original formulation of the Einstein's work \cite{clifton2012,joyce2016,nojiri2017,dodelson2020modern,arbey2021}. Those dark components are identified as dark energy, driving accelerated expansion and dark matter, an entity clustering in galaxies as 'missing mass' needed to properly explain large-scale cosmic structures \cite{dodelson2020modern}. However, current $\Lambda CDM$ model that combines (cold-) dark matter and cosmological constant $\Lambda$ (dark energy) within formulation of GR, undergoes serious pressure, facing new challenges \cite{arbey2021,perivolaropoulos2022}. In addition, many scientists are unsatisfied with the reasoning behind dark sector, since if dark sector does exist, only $5\%$ of matter-energy in the Universe is visible, and dark components often require exotic and strange properties \cite{clifton2012,joyce2015}. In this context, thought that GR may not be the correct proposal to describe gravity in large-scale structure led to the increased interest in alternatives to the current $\Lambda CDM$ description. In such context, modifications of Einstein's formulation are proposed as a different way of explaining cosmic phenomena and experimental issues without invoking dark matter and dark energy \cite{clifton2012,joyce2016}. It is important to note, that some problems may  raise during introduction of new models. The resulting equations are often very complicated and hard to solve, leading to technical difficulties in obtaining new solutions \cite{clifton2012,nojiri2017,marciu2020}. The resulting issues can be bypassed by employing the so-called reconstruction method, commonly used for various, mainly cosmological models \cite{nojiri2017,sardar2022}. The core of the reconstruction approach is to determine the desired evolution first, and then recover (i.e. reconstruct) the corresponding model of interests \cite{bamba2012,akarsu2014,nojiri2017b,kaczmarek2020,kaczmarek2021,paul2022}. 

Note that, one cannot construct a specific cosmic model and 'call it a day', as there is no guarantee that the resulting theory is physically viable \cite{koyama2016,saridakis2021}. Therefore, it is essential to employ various tests and criteria to distinguish physically realizable models from those that are not \cite{clifton2012}. One of such procedures is deeply rooted in the mathematical formalism behind Einstein's general relativity, by imposing certain behaviour on the energy momentum tensor $T_{\mu\nu}$, in the form of energy conditions \cite{poisson2004}. In the conventional GR, these conditions assist in selecting physically realistic matter content of spacetime. In the context of the modified gravity, the energy conditions can be used to constrain the degrees of freedom of the theory and test the physical viability of cosmological scenarios \cite{capozziello2014,kaczmarek2020}. In order to achieve that goal, every alteration and modification to the original GR can be treated as contribution from an effective energy-momentum tensor \cite{poisson2004,santos2007}. By doing so, Universe is filled with some kind of 'fictious' (geometrical) fluids, that in principle can play a role of dark matter and dark energy \cite{capozziello2014, koussour2022}. In general, energy conditions in the literature are referred to and categorized based on four distinct criteria \cite{poisson2004,mandal2020}:
\begin{itemize}
  \item The null energy condition (NEC) - guarantees positivity of matter density along lightlike curves,
  \item The weak energy condition (WEC) - states that the matter density is positive along timelike curves,
  \item The dominant energy condition (DEC) - velocity of the energy-momentum flow cannot be greater than the speed of light,
  \item The strong energy condition (SEC) - ensures attractive behaviour of gravity.
\end{itemize}
Those boundaries proved  useful not only in proving BH theorems but also in the context of cosmology \cite{hawking2023,mandal2020}. Noteworthy, violated SEC corresponds to the accelerated expansion of the Universe \cite{mandal2020,de2022comment}. Therefore, energy conditions serve as crucial tools in defining the physical attributes of proposed modified GR. This significance sparked substantial interest in various modified theories of gravity \cite{capozziello2014,nojiri2017,capozziello2018,yousaf2018,mandal2022}.

Among the plethora of different models, one of the most interesting in recent years is based on the old principles of differential geometry - teleparallelism. The idea, considered even by Einstein himself, focusses on the geometry, emphasizing the role of other notions than curvature in its description. In fact, Levi-Civita connection is not the only possibility for transporting vectors and tensors on the manifold \cite{ortin2004,bahamonde2018,bahamonde2023}. From differential geometry, it is known that general affine connection can be separated into the three parts:
$\Gamma^{\lambda}_{\;\;\mu\nu} = \{^{\lambda}_{\;\;\mu\nu}\} + K^{\lambda}_{\;\;\mu\nu} + L^{\lambda}_{\;\;\mu\nu}$,
with the Levi-Civita connection 
($ \{^{\lambda}_{\;\;\mu\nu}\} = \frac{1}{2} g^{\lambda\sigma} \left( \partial_{\mu} g_{\nu\sigma} + \partial_{\nu} g_{\sigma\mu} - \partial_{\sigma} g_{\mu\nu} \right)$),
  contorsion
($ K^\lambda_{\;\;\mu\nu}=\frac{1}{2}g^{\lambda\beta} \Big( T_{\beta\mu\nu}+T_{\mu\beta\nu} + T_{\nu\beta\mu} \Big)$) characterized by torsion $T_{\;\;\mu\nu}^{\gamma} = \Gamma^{\gamma}_{\;\;\mu\nu} - \Gamma_{\;\;\nu\mu}^{\gamma}$ and disformation $ L^\lambda_{\;\;\mu\nu}=\frac{1}{2}g^{\lambda\beta} \left( Q_{\beta\mu\nu} - Q_{\mu\beta\nu} - Q_{\nu\beta\mu} \right)$ based on nonmetricity $Q_{\gamma\mu\nu} = \nabla_{\gamma}g_{\mu\nu}=\partial_\gamma g_{\mu\nu}-\Gamma^\beta_{\;\;\gamma\mu}g_{\beta\nu}-\Gamma^\beta_{\;\;\gamma\nu}g_{\beta\mu}$ \cite{bahamonde2023}. Moreover, those different approaches merge into some form of trinity referred to as 'geometrical trinity', as they are different representations of the same underlying theory ('three sides of the same coin') \cite{beltran2019}. In this aspect, one of the most intriguing alternative descriptions of gravity is based on flat (curvatureless) connection and vanishing torsion, i.e. $\Gamma^{\lambda}_{\;\; \mu\nu}=L^\lambda_{\;\;\mu\nu}$, being fully determined by nonmetricity tensor \cite{jimenez2018,bahamonde2023}. Theory of that kind is referred to as symmetric teleparallel equivalent of GR (STEGR), where nonmetricity scalar ($Q$) plays a role that usually is performed by Ricci curvature ($R$) \cite{jimenez2018}. The story does not end here, as possible more general forms of nonmetricity theories are grouped in the $f(Q)$ formulation, where gravity action is described by general and analytical function of nonmetricity \cite{jimenez2018,jimenez2020cosmology,jarv2018}. This novel idea opens new possibilities and uncharted territories in the study of the GR alternatives, in a manner similar to the impact of the $f(R)$ theories on the curvature-based modified gravity \cite{sotiriou2006,capozziello2007,clifton2012,nojiri2017}. Note that equations of motion from $f(R)$ gravity are higher order, while remarkably $f(Q)$-theories lead to second order field equations, avoiding Ostrogardski's instabilities \cite{jimenez2018,jarv2018,barros2020,hu2022}.  Currently, $f(Q)$ gravity achieved significant results despite being a few-years old \cite{jimenez2018,mandal2020,gakis2020,frusciante2021,albuquerque2022}. For example, it challenged the aforementioned $\Lambda CDM$ model and provided an alternative that matches with observations \cite{beltran2019,atayde2021,anagnostopoulos2021}. Thus, in the current state-of-the-art, nonmetricity have been studied from different perspectives ranging from the astrophysics and black holes \cite{lazkoz2019,bahamonde2022,calza2022,errehymy2022,mandal2022,bhar2023,calza2023,mishra2023}, to the applications in the theoretical cosmology \cite{bajardi2020,solanki2021,d2022black,koussour2022,arora2022,junior2023,paliathanasis2023,subramaniam2023} and often relying on reconstruction scheme \cite{capozziello2022,myrzakulov2022a,singha2023}. Moreover, energy conditions have been proven useful in obtaining $f(Q)$ models  compatible with accelerated expansion of the Universe \cite{mandal2020,hu2022,mandal2022}. For readers seeking more details, a comprehensive and pedagogical introduction to $f(Q)$ gravity and related 'geometric trinity' is available in \cite{heisenberg2023}.

Moreover, the growing interest in the generalized STEGR (i.e. $f(Q)$ gravity) led to the various extensions and modifications of that theory, by looking for additional degrees of freedom \cite{harko2018}. For example, the nontrivial couplings between matter and nonmetricity tensor have been considered in $f(Q,T)$ \cite{xu2019f} and $f(Q,L_m)$ gravity \cite{harko2018}, for trace of energy momentum tensor $T$ and matter Lagrangian, respectively. Noteworthy, approaches of that kind can provide an alternative to the dark energy \cite{arora2020f,arora2021} or provide viable bounce cosmologies \cite{agrawal2021}.  From slightly different perspective, scalar tensor realizations of $f(Q)$ gravity or metric-affine generalizations (i.e. curvature and torsion is not-vanishing) have also been implemented in different contexts \cite{jarv2018,runkla2018,iosifidis2021,bahamonde2023}. It is important to notice, that there are other different strategies for changing GR. Instead of adding new terms to the action, one can focus on the degrees of freedom that may not be visible upon the first look \cite{sebastiani2017,nojiri2017}. This motivated Chamseddine and Mukhanov to isolate the conformal degree of freedom for the metric, by introducing the parametrization of the physical metric $g$ in terms of scalar field $\phi$ and auxiliary metric $\hat{g}$ in the following way $g_{\mu\nu}=\hat{g}_{\mu\nu}(g^{\alpha\beta}\partial_\alpha \phi \partial_\beta \phi)\equiv P\hat{g}_{\mu\nu}$. It was shown, that utilization of parametrization of that kind leads to the inclusion of the scalar field that 'mimics' the behaviour of the dark matter \cite{chamseddine2013,chamseddine2014y,mirza2019}. Similarly to the theories build upon the nonmetricity scalar, the mimetic theories with various extensions gained significant interest over the recent years, providing another viable GR alternative in the context of dark sector or inflation \cite{nojiri2014m,Astashenok2015,myrzakulov2015,nojiri2016,nojiri2017b,casalino2018,mansoori2021,kaczmarek2021}. However, while torsion-mimetic-teleparallel $f(\mathcal{T})$ gravity has seen some attention, the incorporation of the mimetic degree of freedom into the broader scope of teleparallelism and nonmetricity is still an underexplored area \cite{mirza2019,guo2020,izaurieta2020}. Consequently, the general mimetic-STEGR formulation remains uncharted territory, representing a key task of our current work.

In the present paper, we follow the recent trend in the STEGR with extensions and introduce mimetic generalization of the $f(Q)$ gravity. This is motivated not only by the interesting and desired properties of both mimetic and STEGR theories but also their flexibility in different physically crucial contexts \cite{Astashenok2015,odintsov2015a,odintsov2016,jimenez2018,koussour2023,heisenberg2023}. In details, after brief overview of main characteristic of the STEGR and mimetic approaches, we introduce the corresponding field equations and the main properties of the new theory. The mimetic-$f(Q)$ theory introduced in sec. II can be interpreted as extended STEGR  that is supplied by the conformal degree of freedom arising from the collisionless perfect fluid \cite{chamseddine2013,sebastiani2017}. In section III, we utilize a reconstruction scheme to derive two models capable of describing FLRW cosmology at large scales by employing the hybrid cosmic evolution \cite{akarsu2014,koussour2022}. This choice of scale factor allows us to obtain models suitable for description of the different stages of the cosmic evolution in the concise and unified manner \cite{akarsu2014,nojiri2017}. Firstly, we reconstruct mimetic Lagrange multiplier and potential for a given $f(Q)$ cosmology. In the similar manner, we use inverted reconstruction scheme to obtain mimetic potential and $f(Q)$ functional, that corresponds to the assumed Lagrange multiplier. In section IV, the dynamical behaviour and energy constraints are analysed in terms of redshift, to validate the reconstructed mimetic-$f(Q)$ model for $f(Q)=Q-6\lambda M^2\big(\frac{Q}{6M^2}\big)^\alpha$. By studying those properties, we may verify whether the reconstructed model is viable or not, as realistic behaviour is necessary for proper description of the current Universe' expansion \cite{arbey2021}. In the last part, we present the outline and conclusion of our work, as well as presenting future perspectives that arise from the mimetic approach within generalized mimetic-STEGR. 

\section{Mimetic-$f(Q)$ Framework}
In the following work, we concentrate on mimetic extension of theories built upon nonmetricity alone. Hence, we consider torsionless and curvature free geometry with connection \cite{jimenez2018,koussour2022,heisenberg2023}:
\begin{align}
    \Gamma^{\lambda}_{\;\; \mu\nu}=L^\lambda_{\;\;\mu\nu}=\frac{1}{2}g^{\lambda\beta} \left( Q_{\beta\mu\nu} - Q_{\mu\beta\nu} - Q_{\nu\beta\mu} \right),
    \label{eq1}
\end{align}
characterized solely by the nonmetricity tensor:
\begin{align}
    Q_{\gamma\mu\nu} = \nabla_{\gamma}g_{\mu\nu}=\partial_\gamma g_{\mu\nu}-\Gamma^\beta_{\;\;\gamma\mu}g_{\beta\nu}-\Gamma^\beta_{\;\;\gamma\nu}g_{\beta\mu}.
        \label{eq2}
\end{align}
Thus, geometry build from nonmetricity is defined solely by changes of vectors length upon parallelly transporting them on the manifold \cite{beltran2019}. It is also important to add, that in the STEGR theories the connection can be made vanishing by the appropriate coordinate transformation in the form of so-called coincidental gauge \cite{jimenez2018,beltran2019,heisenberg2023}.

Before  we go deeper into the formulation of theoretical framework, one should note that nonmetricity has two independent traces:
\begin{align}
    Q_{\mu}=Q_{\mu\;\;\nu}^{\;\;\nu},\;\;\;\; \Tilde{Q}_\mu=Q^{\nu}_{\;\;\mu\nu}.
        \label{eq3}
\end{align}
Moreover, one can introduce and define superpotential (conjugate) tensor in terms of nonmetricity and its traces:
\begin{align}
    P^{\alpha}_{\;\;\mu\nu}=\frac{1}{2}Q_{(\mu\;\;\nu)}^{\;\;\alpha}-\frac{1}{4}\Big(Q^{\alpha}_{\;\;\mu\nu}-Q^{\alpha}g_{\mu\nu}-\Tilde{Q}^\alpha g_{\mu\nu}-\delta^\alpha_{(\mu}Q_{\nu)}\Big)
    \label{eq4}
\end{align}
Therefore, nonmetricity scalar can be defined by:
\begin{align}
    Q=-Q_{\alpha\mu\nu}P^{\alpha\mu\nu},
    \label{eq5}
\end{align}
and should be considered as a scalar trace of the nonmetricity tensor (i.e. no free indices left). In what follows, we adopt the coicident gauge, where $\nabla_\mu=\partial_\mu$ and $Q_{\alpha\mu\nu}=\partial_\alpha g_{\mu\nu}$ \cite{jimenez2018,beltran2019}.

On the other hand, in the mimetic gravity, the Einstein-Hilbert metric $g_{\mu\nu}$ can be written in terms of the auxiliary metric tensor $\hat{g}_{\mu\nu}$ and the auxiliary scalar field $\phi$ \cite{chamseddine2013,sebastiani2017}: 
\begin{align}
    g_{\mu\nu}=-\hat{g}^{\alpha\beta}\partial_{\alpha}\phi\partial_{\beta}\phi \hat{g}_{\mu\nu}.
    \label{eq6}
\end{align}
In this manner, the original metric $g_{\mu\nu}$ can be re-expressed with the new degree of freedom linked to the invariance under conformal transformations of the metric $\hat{g}_{\mu\nu}$. Consistency of the Eq.(\ref{eq1}) leads to the mimetic constraint equation imposed on the gradient of the scalar field $\phi$ \cite{chamseddine2013}:
\begin{align}
    g^{\mu\nu}(\phi,\hat{g}_{\mu\nu})\partial_{\mu}\phi\partial_{\nu}\phi=-1.
        \label{eq7}
\end{align}
By employing the mimetic framework, the resulting action can be seen as conformally extended one and Weyl-invariant in terms of metric $\hat{g}_{\mu\nu}$. For the detailed introduction to the mimetic theory, please see \cite{sebastiani2017}.

As a way to impose given constraints, we use well-known Langrange multiplier formalism, extensively used in the mimetic  theories and extensions \cite{odintsov2015a,nojiri2016,mansoori2021,odintsov2016,cardenas2021}. Thus, action of the new mimetic-$f(Q)$ gravity will take the following form:
\begin{align}
S=\int \text{d}^4x\sqrt{-g}\frac{1}{2}\Big[f(Q)-U(\phi)+\eta(g^{\mu\nu}\partial_\mu\phi\partial_\nu\phi+1)  \Big]+\int\text{d}^4 x \sqrt{-g}\mathcal{L}_m
\label{eq8}
\end{align}
with the mimetic field $\phi$ and the scalar (mimetic) potential $U(\phi)$. We note that the presence of the potential in the action serves a minimal extension of the original mimetic framework \cite{chamseddine2014y}. This augmentation allows for the portrayal of a wide array of models,  since any background spacetime can be obtained from apropiate potential $U(\phi)$ \cite{chamseddine2014y}. We also note that the actions involving scalar field constrained by Lagrange multiplier can be studied without referring to the mimetic theory, see for instance \cite{capoziello2010,Gao2011,nojiri2017b,nojiri2019b}. For physical interpretation of the mimetic-$f(Q)$ framework introduced herein, the mimetic degree of freedom (\ref{eq6}) arises from the collisionless perfect fluid, in the spirit of original considerations of Chamseddine and Mukhanov \cite{chamseddine2013,chamseddine2014y}.


Variation of the given mimetic-$f(Q)$ action in terms of $g$ yields the following field equations:
\begin{align}
    \frac{2}{\sqrt{-g}}\nabla_\alpha(\sqrt{-g}f_Q P^\alpha_{\;\;\mu\nu})+\frac{1}{2}g_{\mu\nu}\Big(f(Q)-U(\phi)+\eta(\partial_\alpha \phi \partial^\alpha \phi+1)\Big)-\eta\partial_\mu\phi\partial_\nu \phi+f_Q \big(P_{\mu\alpha\beta}Q_\nu^{\;\;\alpha\beta}-2Q_{\alpha\beta\mu}P^{\alpha\beta}_{\;\;\;\;\nu}\big)=T_{\mu\nu},
    \label{eq9}
\end{align}
or equivalently \cite{heisenberg2023}:
\begin{align}
    f_Q G_{\mu\nu}-\frac{1}{2}g_{\mu\nu}\Big(f(Q)-U(\phi)+\eta(\partial_\alpha \phi \partial^\alpha \phi+1)-Qf_Q\Big)+\eta \partial_\mu \phi \partial_\nu \phi +2f_{QQ}P^\alpha_{\;\;\mu\nu}\partial_\alpha Q=T_{\mu\nu}.
    \label{eq10}
\end{align}
In the above, the energy-momentum tensor is defined as:
\begin{align}
    T_{\mu\nu}=\frac{-2}{\sqrt{-g}}\frac{\delta(\sqrt{-g}\mathcal{L}_m)}{\delta g^{\mu\nu}}.
        \label{eq11}
\end{align}
On the other hand, variation with respect to the connection leads to the:
\begin{align}
    \nabla_\mu\nabla_\nu(\sqrt{-g}P^{\mu\nu}_{\;\;\;\;\alpha})=0.
        \label{eq12}
\end{align}

The equations (\ref{eq9}) or (\ref{eq10}) for $f(Q)=Q$ and $\eta=U(\phi)=0$ reproduce standard Einstein field equations of GR \cite{jimenez2018,beltran2019}. Note the sign convention in the definition of action (\ref{eq8}). In literature, there exist another possibility for extended STEGR with the extra minus sign, for which $f(Q)=-Q$ restores Einstein's formulation \cite{myrzakulov2022a,koussour2022,heisenberg2023}.

Additionally, variation of the action with respect to the mimetic field $\phi$ results in the following scalar equation:
\begin{align}
    2\nabla^{\mu}(\eta \partial_{\mu}\phi)+U'(\phi)=0,
        \label{eq13}
\end{align}
with $U'(\phi)=\frac{\text{d}U(\phi)}{\text{d}\phi}$) and $f_Q=\frac{d Q}{dt}$. It is important to remark, that variation of the action with respect to Lagrange multiplier  $\eta$ returns  mimetic constraint \cite{chamseddine2014y,sebastiani2017}.

Note that, from trace of Eq.(\ref{eq10}) one can get:
\begin{align}
    \eta = f_Q(G-2Q)-2U(\phi)+2f(Q)+2f_{QQ}P^{\alpha\;\;\mu}_{\;\;\mu} \partial_\alpha Q-T,
    \label{eq14}
\end{align}
where we have invoked relationship (\ref{eq7}). Then, inserting it into the Eq.(\ref{eq10}) gives:
\begin{align}\nonumber
        &f_Q G_{\mu\nu}-\frac{1}{2}g_{\mu\nu}\Big(f(Q)-U(\phi)+2f_{QQ}P^\alpha_{\;\;\mu\nu}\partial_\alpha Q +\eta(\partial_\alpha \phi \partial^\alpha \phi+1)-Qf_Q\Big)\\&+\Big (f_Q(G-2Q)-2U(\phi)+2f(Q)+2f_{QQ}P^{\alpha\;\;\mu}_{\;\;\mu} \partial_\alpha Q-T\Big) \partial_\mu \phi \partial_\nu \phi=T_{\mu\nu}.
        \label{eq15}
\end{align}
In that manner we have obtained equations that arise from the equivalent action:
\begin{align}
  S=\int \text{d}^4x\sqrt{-g(\phi,\hat{g}_{\mu\nu})}\frac{1}{2}\Big[f(Q(\phi,\hat{g}_{\mu\nu}))-U(\phi
  )\Big]+\int\text{d}^4 x \sqrt{-g(\phi,\hat{g}_{\mu\nu})}\mathcal{L}_m,  
\label{eq16}
\end{align}
where variation of metric $\delta g_{\mu\nu}$ is treated as variation of  scalar field $\delta \phi$ and auxiliary metric $\delta\hat{g}_{\mu\nu}$, in accordance with the Eq.(\ref{eq6}) and original formulation of Chamseddine and Mukhanov \cite{chamseddine2013,chamseddine2014y,sebastiani2017,mirza2019}. By doing so, we have linked two equivalent formulations of the mimetic-$f(Q)$ gravity. Namely, the action (\ref{eq8}) in terms of $g_{\mu\nu}$ with Lagrange multiplier and action (\ref{eq16}) written in terms of metric $\hat{g}_{\mu\nu}$.

Continuing our work, we will assume that geometry is described by the isotropic and homogeneous FLRW spacetime. The line element for this spacetime takes the following form:
\begin{align}
    \text{d}s^2=-\text{d}t^2+a^2(t) \text{d}x_i\text{d}x^i =-\text{d}t^2+a^2(t)(\text{d}x^2+\text{d}y^2+\text{d}z^2)
    \label{eq17}
\end{align}
in the usual Cartesian coordinates $(t,x,y,z)$ \cite{dodelson2020modern}. We also assume that scalar field depends only on time, {\it i.e.} $\phi=\phi(t)$. 
Then, the nonmetricity in coincidence gauge will take the simple form \cite{jimenez2020cosmology,koussour2022}:
\begin{align}
    Q=6H^2,
    \label{eq18}
\end{align}
where 'dot' ($\dot{}$) denotes differentiation w.r.t. cosmic time $t$ and Hubble rate is given by $H=\frac{\dot{a}}{a}$.
In addition, we assume perfect fluid matter content of the Universe, described by the energy-momentum tensor \cite{hobson2006}:
\begin{align}
    T_{\mu\nu}=(\rho+p)v_\mu v_\nu+p g_{\mu\nu},
    \label{eq19}
\end{align}
for the four-velocity of the fluid $v_\mu$ that satisfy $v^\mu v_\mu=-1$. The energy density and pressure are denoted as $\rho$ and $p$, respectively.
Furthermore, we can assume that perfect fluid satisfies standard isotropic equation of state (EoS):
\begin{align}
    w\rho=p,
    \label{eq20}
\end{align}
characterized by parameter $w$. Additionally,  vanishing four-divergence of the energy-momentum ($\nabla^\mu T_{\mu\nu}=0$) leads to the conservation equation:
\begin{align}
    \dot{\rho}+3H\rho(1+w)=0,
    \label{eq21}
\end{align}
leading to the general form of energy density for cosmic fluid:
\begin{align}
    \rho(t)=\rho_0 \Big(\frac{a}{a_0}\Big)^{-3(w+1)}.
    \label{eq22}
\end{align}
Then, corresponding nonmetricity field equations are:
\begin{align}
    6H^2 f_Q-\frac{1}{2}\Big(f(Q)-U(\phi)+\eta(\dot{\phi}^2+1)\Big)=\rho
    \label{eq23}
\end{align}
and
\begin{align}
    2(\dot{H}+3H^2)f_Q+2H\dot{f}_Q-\frac{1}{2}(f(Q)-U(\phi))=-p,
    \label{eq24}
\end{align}
for $00$th and $ii$th components respectively.
Moreover, the scalar equation take form:
\begin{align}
    6H\eta \dot{\phi}+2(\dot{\eta}\dot{\phi}+\eta\Ddot{\phi})-U'(\phi)=0.
\label{eq25}
\end{align}
Mimetic constraint for the FLRW metric leads to the association of the scalar field with a cosmic time $\phi=t$. We note that, this is an important result in the mimetic gravities as in a wide class of the scalar-tensor theories, the scalar field coupled to gravity often takes more complicated forms.
By using $\phi=t$, the field and scalar equations (\ref{eq23}-\ref{eq25}) reduce to:
\begin{align}
    6H^2 f_Q-\frac{1}{2}\big(f(Q)-U(t)+2\eta\big)=\rho,
    \label{eq26} \\
    2(\dot{H}+3H^2)f_Q+H\dot{f}_Q-\frac{1}{2}\big(f(Q)-U(t)\big)=-p,
    \label{eq27}\\
    6H\eta +2\dot{\eta}-\dot{U}=0.
\label{eq28}
\end{align}
Combining equations (\ref{eq26}) and (\ref{eq27}) results in the expression for Lagrange multiplier:
\begin{align}
    \eta(t)=-\rho-p-H\dot{f}_Q-2f_Q \dot{H}.
    \label{eq29}
\end{align}
Note that for $f(Q)=Q$, standard mimetic gravity in Lagrange multiplier formalism is obtained in Eqs. (\ref{eq21}-\ref{eq24}) \cite{sebastiani2017}. This confirms that the construction of the mimetic-$f(Q)$ formalism is done properly, since in the exact same manner $f(Q)=Q$ in standard $f(Q)$ gravity recovers equations of GR: $3H^2=\rho$ and $2\dot{H}+3H^2=-p$ \cite{jimenez2018,mussatayeva2023}.
For the particular form of $f(Q)$ gravity, by solving Eq.(\ref{eq28}), one can obtain the desired potential. Additionaly, the Lagrange multiplier can be found from Eq.(\ref{eq29}). This scheme will be covered in details in the next section.
\section{Reconstruction of the Hybrid Expansion Law}
Obtaining new solutions for field equations is not an easy task, due to the inherent complexity and nonlinearity. Hence, studying cosmic scenarios is hard to tackle as there is no guarantee that new model is solvable. That troubles are even more evident and hard to bypass where GR modifications play a role  \cite{clifton2012}. However, in the reconstruction procedure, the field equations are inverted to unveil the class of modified theories that give rise to a specific flat Friedmann-Lema{\^ i}tre-Robertson-Walker (FLRW) model. In other words, instead of solving field equations for a given model in order to obtain scale factor, the equations are used to reconstruct a model that satisfy a given scale factor \cite{odintsov2015a,odintsov2016,kaczmarek2021,gadbail2022,gonccalves2022b}. It is obvious, that ability to describe well-established cosmological models in the GR modifications is desired \cite{joyce2016,nojiri2017}.
The hybrid expansion law (HEL) was introduced in \cite{akarsu2014} as a product of the power-law evolutions and the flat de-Sitter cosmology. Besides describing these two phases, the hybrid scale factor allows for elegant and nice transition between those stages in cosmic history, namely deceleration to the cosmic acceleration \cite{akarsu2014}. This is particularly interesting property for reconstruction procedure in the modified gravity, as one can at hand reconstruct a model that describes cosmic history in the unified manner \cite{nojiri2017}. The corresponding scale factor and Hubble rate take the following form \cite{akarsu2014,koussour2023}:
\begin{align}
    a(t)=a_0 \Big(\frac{t}{t_0}\Big)^{A}e^{B\big(\frac{t}{t_0}-1\big)},\;\;\; H=\frac{A}{t}+\frac{B}{t_0},
\label{eq30}
\end{align}
where $A\geq 0$ and $B\geq 0$ and the present values of the scale factor and Universe' age are $a_0$ and $t_0$ accordingly \cite{akarsu2014}.

\subsection{Reconstruction for $\eta(t)$ and $U(t)$}
In this part, we will focus on obtaining Lagrange multiplier and potential for the corresponding $f(Q)$ gravity model. For this purpose, we have to specify the functional form of $f(Q)$. In this manner we will follow the steps of Jimenez, Heisenberg and Koivisto in their pioneering work, by considering the nonmetricity functional of \cite{jimenez2018}:
\begin{align}
f(Q)=Q-6\lambda M^2\big(\frac{Q}{6M^2}\big)^\alpha,
\label{eq31}
\end{align}
where $M$ should be treated as some scale, $\lambda$ is a dimensionless parameter of order $1$. For the functional specified above, $\alpha=1/2$ recovers usual mimetic characteristics within standard GR for $M^2>>H^2$, while modifications appear when $M^2\lesssim H^2$ \cite{jimenez2018}.
Thus, the combined equation (\ref{eq29}) for $\eta$ is:
\begin{align}
    \frac{2 \alpha ^2 \dot{H} \lambda  M^2 \left(\frac{H^2}{M^2}\right)^{\alpha }-H^2 \left(\eta +2
   \dot{H}+p+\rho \right)}{H^2}=0.
   \label{eq32}
\end{align}
Combining the Lgrange multiplier from (\ref{eq32}) with the Klein-Gordon equation (\ref{eq28}), leads to:
\begin{align}\nonumber
 H^3\frac{d U}{dt}&= 4 \alpha  (2 \alpha -1) \lambda  M^2 H \ddot{H}
   \left(\frac{H^2}{M^2}\right)^{\alpha }-2 H^3 \left(2 \ddot{H}+\dot{p}+\dot{\rho}\right)+8 \alpha 
   \left(2 \alpha ^2-3 \alpha +1\right) \lambda  M^2 \dot{H}^2 \left(\frac{H^2}{M^2}\right)^{\alpha }\\  &+12
   \alpha  (2 \alpha -1) \lambda  M^4 \dot{H} \left(\frac{H^2}{M^2}\right)^{\alpha +1}-6 H^4 \left(2
   \dot{H}+p+\rho\right).
\label{eq33}
\end{align}
Thus, once the Lagrange multiplier is obtained, one can solve the differential equation with respect to the mimetic potential $U$. Now, from applying scale factor (\ref{eq30}) in Eqs. (\ref{eq32}) and (\ref{eq33}) and keeping in mind the energy density (\ref{eq22}),  the reconstructed Lagrange multiplier and potential are:
\begin{align}\nonumber
    \eta(t)&=\frac{2 A \left(\text{t0}^2 \left(A^2-\alpha  (2 \alpha -1) \lambda  M^2 t^2 \left(\frac{(A \text{t0}+B
   t)^2}{M^2 t^2 \text{t0}^2}\right)^{\alpha }\right)+2 A B t \text{t0}+B^2 t^2\right)}{t^2 (A
   \text{t0}+B t)^2}\\&+\text{$\rho $0} (w+1) \left(\frac{t}{\text{t0}}\right)^{-3 A} \left(-e^{B \left(3-\frac{3
   t}{\text{t0}}\right)}\right) \left(\left(\frac{t}{\text{t0}}\right)^A e^{B
   \left(\frac{t}{\text{t0}}-1\right)}\right)^{-3 w}
   \label{eq34}
\end{align}
and
\begin{align}\nonumber
U(t)&=\frac{2 (2 \alpha -1) \lambda  \left(6 A B t \text{t0}+A \text{t0}^2 (3 A-2 \alpha
   )+3 B^2 t^2\right) \left(\frac{(A \text{t0}+B t)^2}{M^2 t^2 \text{t0}^2}\right)^{\alpha -1}}{t^2
   \text{t0}^2}-\frac{2 A ((3 A-2) \text{t0}+6 B t)}{t^2 \text{t0}}\\ &-2 \text{$\rho $0} w
   \left(\frac{t}{\text{t0}}\right)^{-3 A} e^{B \left(3-\frac{3 t}{\text{t0}}\right)}
   \left(\left(\frac{t}{\text{t0}}\right)^A e^{B \left(\frac{t}{\text{t0}}-1\right)}\right)^{-3
   w}+U_0.
      \label{eq35}
\end{align}
Thus, it is possible to obtain hybrid-evolution law scale factor in the mimetic framework of $f(Q)$, once the specific functional for extended STEGR is specified. Moreover for the vacuum ($\rho=p=0$), in the limit of $t\rightarrow \infty$ (i.e. $A=0$) corresponding to the de-Sitter evolution, the resulting potential is constant $U(t)=6 (2 \alpha -1) \lambda  M^2 \left(\frac{B^2}{M^2 t_0^2}\right)^{\alpha }+U_0$, while $\eta$ vanishes. Thus, for the late time cosmic acceleration, the constant potential will be responsible, mimicking the dark energy \cite{sebastiani2017}. We note that this is a common characteristic for mimetic extensions of GR modifications and alternatives \cite{odintsov2016,mirza2019,sebastiani2017,kaczmarek2021,zhang2023}.

\subsection{Reconstruction for $f(Q)$ and $U$ from the Lagrange multiplier}

In this scenario, instead of specifying a functional form of nonmetricity $f(Q$ we will focus on obtaining a model that describes HEL for a specified Lagrange multiplier. Herein we present the general procedure first, after which we will use it for particular choice of $\eta$. In what follows, once the multiplier $\eta$ is known, the Eq.(\ref{eq29}) can be solved for $f_Q (t)$, instead of $\eta$. In fact, the resulting differential equation can be easily solved by integration with respect to the time $t$:
\begin{align}
    f_Q(t)=\frac{1}{H^2}\Big[\int-\Big(\rho(t)+p(t)+\eta(t)\Big)H\text{d}t +f_0\Big],
        \label{eq36}
\end{align}
where $f_0$ is the integration constant. Using the relationship between nonmetricity scalar and Hubble rate $Q=6H^2$ (i.e. Eq.(\ref{eq18})) one can get an explicit form of  $t(Q)$ and obtain a function $f_Q(Q)$. Then, Eq.(\ref{eq36}) can be easily integrated in order to find $f(Q)$:
\begin{align}
    f(Q)=\int f_Q(Q)\text{d}Q.
        \label{eq37}
\end{align}
On the other hand, scalar equation can be solved for $U$, leading to:
\begin{align}
 U(t)=U_0+ \int _1^t2
   \left(3 H(t_1) \eta
   (t_1)+\eta'(t_1)\right)dt_1  
        \label{eq38}.
\end{align}
Note, that the inverted reconstruction is a standard approach in theories built on the mimetic degree of freedom, mainly used to showcase the flexibility of the introduced model \cite{nojiri2014m,Astashenok2015}. Thus, one should treat this approach as a complement to the procedure described in the previous subsection. Similarly to the mimetic $f(R)$ gravity, one can also specify the potential $U(\phi)$ and then obtain the Lagrange multiplier and corresponding action functional \cite{nojiri2014m}.

In what follows, we will now reconstruct the $f(Q)$ functional for specified Lagrange multiplier. The chosen multiplier will have simple quadratic form in terms of Hubble factor:
\begin{align}
    \eta(t)= \eta_0 +\gamma H^2,
    \label{eq39}
\end{align}
Now, due to the complexity of the equations, we will focus on the vacuum case by taking $\rho=w=0$. This is a standard approach in mimetic gravity, as one of the reason for introduction that kind of modification to Einsten's theory is that it can 'mimic' matter \cite{chamseddine2013,sebastiani2017}.
Hence, from $Q=6H^2$, the inverted function $t(Q)$ can be obtained:
\begin{align}
    t= \frac{-6 A B t_0-\sqrt{6} A
   \sqrt{Q} t_0^2}{6 B^2-Q
   t_0^2}\;\;\;\;\text{or}\;\;\;\;t= \frac{\sqrt{6} A
   \sqrt{Q} t_0^2-6 A B t_0}{6 B^2-Q
   t_0^2}.
   \label{eq40}
\end{align}
Since we are dealing with the positive time, we restrict ourselves to the second option of Eq.(\ref{eq40}).
The potential corresponding to the given Lagrange multiplier, can be obtained immediately, by integrating Eq.(\ref{eq38}) and takes form:
\begin{align}
   U(t)=\frac{2 B
   \left(-9 A^2 \gamma +2 A \gamma +3
   \eta_0 t^2\right)}{t
   t_0}+\frac{(2-3 A) A^2
   \gamma }{t^2}+\frac{6 A \log (t)
   \left(3 B^2 \gamma +\text{$\eta
   $0}
   t_0^2\right)}{t_0^2}+\frac{6 B^3 \gamma 
   t}{t_0^3}+U_0.
   \label{eq41}
\end{align}
On the other hand, from Eq.(\ref{eq36}) one get the following form of $f_Q(t)$:
\begin{align}
    f_Q(t)= \frac{A^3 \gamma  t_0^3-2 B t t_0^2
   \left(\eta_0 t^2-3 A^2 \gamma \right)-2 A t^2 t_0 \log (t) \left(3 B^2
   \gamma +\eta_0 t_0^2\right)-2 B^3 \gamma  t^3+2 c_1 t^2 t_0}{2
   t_0 (A t_0+B t)^2}.
   \label{eq42}
\end{align}
Now, inserting the relation for $t(Q)$ of (\ref{eq40}) leads to:
\begin{align}\nonumber
    f_Q(Q)&=\frac{t_0^2 \left(\sqrt{6} \sqrt{Q} t_0-6 B\right)^2}{\left(\sqrt{6} B \sqrt{Q} t_0^2-Q
   t_0^3\right)^2}\Big[ -A \left(3 B^2 \gamma +\eta_0 t_0^2\right) \log \left(\frac{A t_0
   \left(\sqrt{6} \sqrt{Q} t_0-6 B\right)}{6 B^2-Q t_0^2}\right)-\frac{3 A B
   \gamma  \left(6 B^2-Q t_0^2\right)}{6 B-\sqrt{6} \sqrt{Q} t_0}\\ &+\frac{A
   \gamma  \left(Q t_0^2-6 B^2\right)^2}{2 \left(\sqrt{6} \sqrt{Q} t_0-6
   B\right)^2}+\frac{A B \eta_0 t_0^2 \left(6 B-\sqrt{6} \sqrt{Q}
   t_0\right)}{6 B^2-Q t_0^2}+\frac{A B^3 \gamma  \left(6 B-\sqrt{6} \sqrt{Q}
   t_0\right)}{6 B^2-Q t_0^2}+c_1  \Big].
\label{eq43}
\end{align}
This lengthy expression can be integrated with respect to the $Q$, leading to the explicit form of $f(Q)$ functional in the mimetic-nonmetricity regime introduced herein. For the full form of the solution of Eq.(\ref{eq43}), we refer a reader to the Appendix.
Thus, even the mathematically complicated forms of scale factor can be obtained within mimetic-$f(Q)$, showing great flexibility and adaptation of that framework. Noteworthy, the FLRW cosmology has been reconstructed without invoking any matter content and as a consequence, full hybrid cosmic evolution is characterized only by nonmetricity and mimetic constraint from $f(Q)$ and $\eta$ with $U$, respectively. Note that from the practical point of view, the standard approach with the specified $f(Q)$ functional is a more natural one, since the resulting models usually are easier to handle and less mathematically complicated. In fact, this is a common property of the theories twin to the mimetic-$f(Q)$ introduced herein, such as mimetic-$f(R)$ or other higher-order models \cite{nojiri2014m,Astashenok2015,sebastiani2017,kaczmarek2021}.

\section{Dynamical properties}
\subsection{Redshift and equation of state}
\begin{figure}
    \centering
    \includegraphics[scale=0.91]{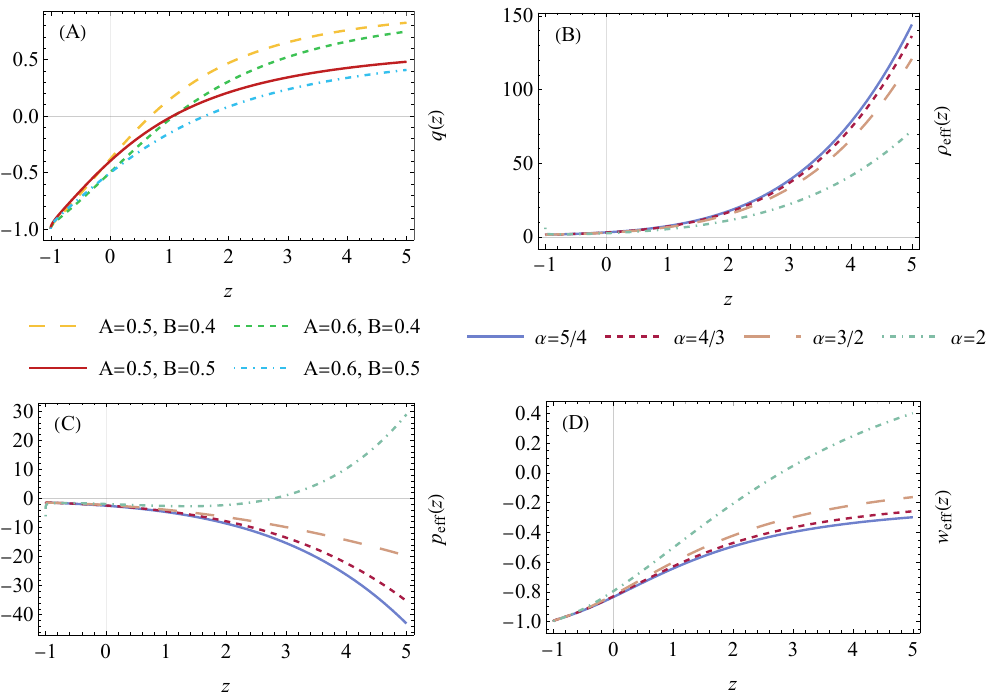}
    \caption{ Evolution of the deceleration  $q(z)$ (A) for different combinations of hybrid scale factor parameters $A$ and $B$ in terms of redshift $z$. Visualization of the energy density (B), effective pressure (C) and EoS parameter (D) as a function of $z$ for different values of $\alpha$,where $A=B=0.5$.}
    \label{fig1}
\end{figure}

In this section, we will focus on main properties of the model reconstructed in (A) part of the section $II$. In order to better relate to the current observations and data, it is often more advantageous  to use cosmic redshift $z$ instead of  time $t$.  Redshift $z$ usually serves as a natural cosmic measure, revealing not just distance but also the temporal context of celestial phenomena. The higher $z$ value is, the corresponding cosmic epoch is older and the distance to the remote object is greater \cite{dodelson2020modern}. For example, the beginning of the universe corresponds to the infinite redshift ($z \rightarrow \infty$). As redshift is directly associated with the geometry, the relation between $z$ and scale factor is:
\begin{align}
    a(t)=\frac{a_0}{1+z},
    \label{eq44}
\end{align}
for the present value $a_0$ of a scale factor. In the case of the hybrid expansion of (\ref{eq30}), the age of Universe as a function of redshift reads \cite{koussour2023}:
\begin{align}
    t(z)=\frac{A t_0}{B}f(z),
    \label{eq45}
\end{align}
for $f(z)=W[\frac{B}{A} \exp(\frac{B-\ln(1+z)}{A})]$, where $W$ is a Lambert function ('product logarithm') \cite{corless1996,koussour2023}.
Moreover, Hubble factor and its time derivative in terms of redshift $z$ are accordingly:
\begin{align}
    H(z)=\frac{H_0 B}{B+A}[1+\frac{1}{f(z)}]
    \label{eq46}
\end{align}
and 
\begin{align}
    \dot{H}=-(1+z)H(z)\frac{d H(z)}{d z},
    \label{eq47}
\end{align}
where $H_0=\frac{A+B}{t_0}$ stands for current value of Hubble parameter. Moreover,  the deccelaration parameter $q(z)$:
\begin{align}
q(z) = \frac{1}{H(z)^2} \left[ (1 + z) \frac{dH(z)}{dz} - 1 \right]
\label{eq48}
\end{align}
can be introduced to address dynamics and behaviour of the expansion of Universe. From its dynamics, one can easily deduce whether accelerated expansion is slowing down \cite{dodelson2020modern,koussour2023}. The  deceleration as a function of redshift in the hybrid evolution scenario of (\ref{eq30}) has been plotted on panel (A) of Fig.(\ref{fig1}) for different choices of $A$ and $B$. In the present work, we will focus on the choice $A=B=1/2$. For general details on dynamical behaviour of hybrid expansion law, we refer to \cite{akarsu2014,koussour2023,goswami2019}, where other choices of $A$ and $B$ have been studied.  

Now, we can address the dynamics of the reconstructed model for varying redshift $z$. Firstly, we express the field equations in the form more suitable for studying energy conditions and related properties. In order to achieve this, we rewrite them in the Eisteinian form:
\begin{align}
    3H^2=\rho_{eff},\;\;\;\;-(\dot{H}+3H^2)=p_{eff}
    \label{eq49}
\end{align}
where we have defined the effective energy density as:
\begin{align}
    \rho_{eff}=\frac{1}{2f_Q}\Big[ \rho + \frac{1}{2}\big( f(Q)-U(t)+2\eta  \big) \Big],
    \label{eq50}
\end{align}
and effective pressure in the form of:
\begin{align}
    p_{eff}=\frac{1}{2f_Q}\Big[p+H f_{QQ}\dot{Q}-\frac{1}{2}\big(f(Q)-U(t)\big)\Big].
    \label{eq51}
\end{align}
The reader should treat the above reformulation as follows. Every term that is not the ordinary left-hand side of FLRW equations in GR, can be treated as effective fluid and component of the effective energy-momentum tensor $T_{\mu\nu}^{eff}$ \cite{nojiri2017,mandal2020}. Moreover, any geometrical quantity coming from the mimetic-$f(Q)$ action can be regarded as the dark fluid that contributes to the matter content of the Universe. We note, that this is standard association in modified gravity theories  \cite{clifton2012,nojiri2017}.

Additionally, in a manner similar to the usual matter content of the universe, one can also define the effective equation of state parameter:
\begin{align}
    w_{eff}=\frac{p_{eff}}{\rho_{eff}}=\frac{2[H \dot{f}_Q+p+\eta+\rho]}{f(Q)+2 (\eta+\rho
   )-U(t)}-1.
   \label{eq52}
\end{align}
As a check of consistency of the mimetic extension presented herein, we note that for $f(Q)=Q$ and $\eta=U=0$, perfect fluid EoS of Einstein's formulation is recovered. For the function of $Q$ from Eq(\ref{eq31}), we get:
\begin{align}
    \rho_{eff}(z)=\frac{-3 \lambda  M^2
   \left(\frac{H^2}{M^2}\right)^{\alpha
   }+3 H^2+\eta (t)+\rho
   (t)-\frac{u(t)}{2}}{2 \left(1-\alpha 
   \lambda 
   \left(\frac{H^2}{M^2}\right)^{\alpha
   -1}\right)},
   \label{eq53}
\end{align}
and
\begin{align}
    p_{eff}(z)=\frac{4 (\alpha -1) \alpha  \lambda  M^2 \dot{H} \left(\frac{H^2}{M^2}\right)^{\alpha }-H^2 \left(6 \lambda  M^2
   \left(\frac{H^2}{M^2}\right)^{\alpha }+2 p+U\right)+6 H^4}{4 \alpha  \lambda  M^2
   \left(\frac{H^2}{M^2}\right)^{\alpha }-4 H^2}.
   \label{eq54}
\end{align}
Dynamical nature of the $\rho_{eff}$ and $p_{eff}$ is visualized respectively on the panels (B) and (C) of Fig.(\ref{fig1}), where four different choices of $\lambda$ have been considered. As can be seen, the effective energy density stayed positive in the past. This is not the case for the effective pressure that is currently negative at $z=0$, and except mimetic-$f(Q)$ model with $\alpha=2$, it decreases as the redshift increase. Since negative pressure is linked with the repulsive behaviour of gravity, the reconstructed model may characterize accelerated expansion, since negative pressure is one of the main characteristic of dark energy (cosmological constant) \cite{arbey2021}.
Moreover, the corresponding EoS is given by:
\begin{align}
    w_{eff}(z)=\frac{-4 (\alpha -1) \alpha  \lambda  M^2 \dot{H} \left(\frac{H^2}{M^2}\right)^{\alpha }+H^2 \left(6 \lambda  M^2
   \left(\frac{H^2}{M^2}\right)^{\alpha }+2 p+U\right)-6 H^4}{H^2 \left(2 \left(-3 \lambda  M^2
   \left(\frac{H^2}{M^2}\right)^{\alpha }+\eta+\rho\right)+6 H^2-U\right)}
   \label{eq55}
\end{align}
Having the effective EoS parameter, one can study and categorize behaviour of the dark sector with its (effective) components. Currently, (i.e. $z=0$), the observations  suggest the equation of state parameter to be $w_{eff}<-1/3$, in order to address accelerated expansion of the Universe. The behaviour of the effective $w_{eff}$ for the model of Eq.(\ref{eq31}) is presented on (D) plot of Fig.(\ref{fig1}), where different values of $\alpha$ has been taken. In current times, EoS enters quintessence region of $w_{eff}\in (-1,1/3)$ and approaches $\Lambda CDM$ value $w_{eff}=-1$ in the future as $z\rightarrow -1$. Hence, the model of mimetic $f(Q)$ specified by Eq.(\ref{eq31}) for the scale factor (\ref{eq30}) leads to the large acceleration phase for small redshift, resembling properties described by the recent observations.
\begin{figure}
    \centering
    \includegraphics[scale=0.91]{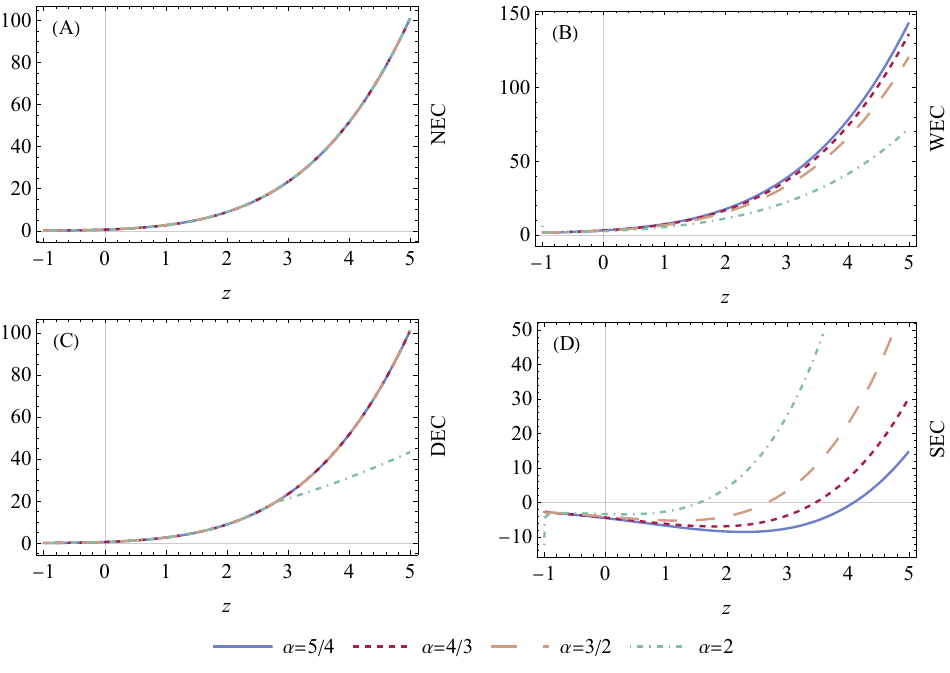}
    \caption{The energy conditions for the mimetic-$f(Q)$ gravity specified by (\ref{eq31}) for different values of parameter $\alpha$, where $M=t_0=1,\lambda=\rho_0=a_0=2,w=C_1=0$. }
    \label{fig2}
\end{figure}

\subsection{Energy conditions}
After studying basic properties of the reconstructed mimetic-$f(Q)=Q-6\lambda M^2\big(\frac{Q}{6M^2}\big)^\alpha$ gravity, one can determine the energy constraints that have to be checked for the effective fluid distribution of the Universe described by HEL. In order to study energy conditions in general relativity, one should start with the Raychaudhury equations:
\cite{poisson2004}:
\begin{align}
\frac{d\theta}{d\tau}=-\frac{1}{3}\theta^2+\omega_{\mu\nu}\omega^{\mu\nu}-\sigma_{\mu\nu}\sigma^{\mu\nu}-R_{\mu\nu}u^\mu u^\nu,\;\;\;
\frac{d\theta}{d\tau}=-\frac{1}{2}\theta^2+\omega_{\mu\nu}\omega^{\mu\nu}-\sigma_{\mu\nu}\sigma^{\mu\nu}-R_{\mu\nu}k^\mu k^\nu.
\label{eq56}
\end{align}
where $\theta$, $\sigma_{\mu\nu}$, and $\omega_{\mu\nu}$ denote respectively expansion factor, shear and rotation. the timelike and null tangent vectors are denoted as $u^{\mu}$ and $k^{\mu}$. Those equations relate matter-geometry and behaviour of congruences of curves in the spacetime manifold. Therefore, for attractive gravity, Raychaudhury equations have to satisfy the presented conditions:
\begin{align}
R_{\mu\nu}u^\mu u^\nu=\Big( T_{\mu\nu}-\frac{1}{2}g_{\mu\nu}T\Big)u^\mu u^\nu\geq 0,\;\;\;\;
R_{\mu\nu}k^\mu k^\nu=\Big( T_{\mu\nu}-\frac{1}{2}g_{\mu\nu}T\Big)k^\mu k^\nu\geq 0.
\label{eq57}
\end{align}
Those geometric constraints (\ref{eq57}) yield energy conditions for a perfect fluid matter distribution:
\begin{itemize}
  \item NEC: $\rho+p\geq 0$,
  \item WEC: $\rho\geq 0$,   $\rho+p\geq 0$, 
  \item SEC: $\rho+p\geq 0$, $\rho+3p\geq 0$,
  \item DEC: $\rho\geq 0$, $\rho\geq |p|$,
\end{itemize}
that obey $
DEC \Longrightarrow WEC \Longrightarrow NEC \Longleftarrow SEC$. Hence, if the NEC is violated, the remaining energy conditions will also be violated \cite{kaczmarek2020}. For the modified theories of gravity, energy constraints can be extended, due to the geometric character of the Raychaudhuri equations. Nevertheless, certain ambiguities might arise, such as alteration of the energy constraints upon transitioning between conformally related frames \cite{capozziello2014}. For a comprehensive discussion concerning energy conditions in modified gravity, we refer to \cite{capozziello2014,capozziello2015,capozziello2018}.

Hence, for the hybrid cosmic evolution \ref{eq41} and mimetic-$f(Q)$ gravity specified by \ref{eq31}, the associated energy conditions are:
\begin{align}
    \text{NEC}: \rho_{eff}+p_{eff}=\frac{H^2 (p+\eta +\rho)-2 (\alpha -1) \alpha  \lambda  M^2 \dot{H}
   \left(\frac{H^2}{M^2}\right)^{\alpha }}{2 \left(H^2-\alpha  \lambda  M^2
   \left(\frac{H^2}{M^2}\right)^{\alpha }\right)}\geq,
   \label{eq58}
\end{align}
\begin{align}
        \text{WEC}: \rho_{eff}(z)=\frac{-3 \lambda  M^2
   \left(\frac{H^2}{M^2}\right)^{\alpha
   }+3 H^2+\eta +\rho
   -\frac{U}{2}}{2 \left(1-\alpha 
   \lambda 
   \left(\frac{H^2}{M^2}\right)^{\alpha
   -1}\right)}\geq 0,
   \label{eq59}
\end{align}
\begin{align}\nonumber
    \text{DEC}: &\rho_{eff}-|p_{eff}|=\frac{-3 \lambda  M^2 \left(\frac{H^2}{M^2}\right)^{\alpha }+3 H^2+\eta +\rho
   -\frac{U}{2}}{2 \left(1-\alpha  \lambda  \left(\frac{H^2}{M^2}\right)^{\alpha
   -1}\right)}\\ &-\bigg| \frac{4 M^2 (\alpha -1) \alpha  \lambda  \dot{H}
   \left(\frac{H^2}{M^2}\right)^{\alpha }+6 H^4-H^2 \left(6 M^2 \lambda 
   \left(\frac{H^2}{M^2}\right)^{\alpha }+2 p+U\right)}{4 M^2 \alpha  \lambda 
   \left(\frac{H^2}{M^2}\right)^{\alpha }-4 H^2}\bigg|\geq 0,
   \label{eq60}
\end{align}

\begin{align}
 \text{SEC}: \rho_{eff}+3p_{eff}=\frac{-6 (\alpha -1) \alpha  \lambda  M^2 \dot{H} \left(\frac{H^2}{M^2}\right)^{\alpha
   }+H^2 \left(6 \lambda  M^2 \left(\frac{H^2}{M^2}\right)^{\alpha }+3 p+\eta
   +\rho +U\right)-6 H^4}{2 \left(H^2-\alpha  \lambda  M^2
   \left(\frac{H^2}{M^2}\right)^{\alpha }\right)}\geq 0.
   \label{eq61}
\end{align}
Evolution of the given energy conditions in terms of redshift is presented on panels (A)-(D) of the Fig.(\ref{fig2}), where different possible values of $\alpha$ have been taken into account. In general, three of the four energy conditions are satisfied, as contribution from effective energy density $\rho_{eff}$ dominates inequalities of NEC and DEC. Then, WEC is naturally satisfied, since $\rho_{eff}$ stays positive, as it was discussed in the previous section. On the other hand, SEC is generally violated in the current ($z=0$) and the future ($z=-1$) stage, leading to the repulsive behaviour of gravity. In such manner, acceleration of the universe can be currently predicted \cite{visser2000,koussour2022}. Hence, the energy conditions for mimetic-$f(Q)$ gravity, with the $f(Q)=Q-6\lambda M^2\big(\frac{Q}{6M^2}\big)^\alpha$ are well-behaved in the context of the cosmic dynamics. The resulting equation of state parameter $w_{eff}$ is dynamical and 'mimics' quintessence \cite{joyce2016}. Additionally, the general behaviour of energy conditions for a broad range of parameters $\alpha$ is presented on the Fig(\ref{figApp}) in the Appendix.
\section{Summary and conclusions}
The century after the Einstein original formulation of GR, there at three equivalent descriptions of the gravity, based on the different construction of the connection in the so-called geometrical trinity of gravity \cite{beltran2019,heisenberg2023}. This intriguing fact led to the possible existence of even more general framework, where formulations based on nonmetricity, torsion or curvature  theories are particular cases. Additionally, can this geometrical trinity extend to the modifications of the GR, merging different ideas and motivations \cite{sebastiani2017}?

Within this context, the following work explores the new ground in the modified GR theories by combines two interesting approaches within a single, unified framework of the mimetic-$f(Q)$ gravity. The description introduced herein can be understood and interpreted in two ways: i) as a mimetic gravity with the gravitational action driven by the functional of the nonmetricity tensor, similarly to the teleparallel mimetic $f(T)$ \cite{chamseddine2013,chamseddine2014y,mirza2019}; ii) as the $f(Q)$ gravity with the degree of freedom introduced by the presence of the presureless fluid \cite{sebastiani2017,jimenez2018,jimenez2020cosmology}. Similarly to the original curvature-based formulation of mimetic gravity, the extended $f(Q)$ approach can be presented in the two equivalent actions (\ref{eq8}) and (\ref{eq16}), leading to the two equivalent sets of equations of motion. Moreover, the obtained equations reduce to the usual mimetic gravity for $f(Q)\rightarrow Q$, proving equivalence between two different constructions of geometry, in accordance with the geometric trinity in gravity \cite{beltran2019,bahamonde2023}. As a next step, we have obtained Friedmann equations within the mimetic-nonmetricity framework. In this regard, extra terms present in the modified $f(Q)$ FLRW equations, can be viewed as a contribution from the additional degree of freedom implemented via Lagrange multiplier. After the discussion of the main properties of the resulting theory, we have used the cosmological reconstruction method to study the hybrid cosmology characterized by scale factor (\ref{eq30}): $a(t)=a_0 \Big(\frac{t}{t_0}\Big)^{A}e^{B\big(\frac{t}{t_0}-1\big)}$. By doing so, the question of finding models that can provide cosmic evolution for distinct phases is answered, as the hybrid law describing two separate stages of the universe history can be achieved, within the framework of interest. Specifically, we have obtained two models from two different approaches to the reconstruction method. In the first scenario, the Lagrange multiplier $\eta(t)$ and potential $U(t)$ have been obtained for the assumed functional of $f(Q)=Q-6\lambda M^2\big(\frac{Q}{6M^2}\big)^\alpha$.  In the second case, using the inverse reconstruction method for known Lagrange multiplier $\eta_0+\gamma H^2$, we have obtained the corresponding $f(Q)$ function and potential $U(\phi)$ in the absence of the matter fields. Note, that considerations of that type are an inherent feature of the mimetic gravity, as it 'mimic' evolution usually driven by some matter fields \cite{chamseddine2013}. Then, we have studied the dynamics for the model with $f(Q)=Q-6\lambda M^2\big(\frac{Q}{6M^2}\big)^\alpha$ in terms of the redshift $z$. After analysing behaviour of the hybrid model for different choices of constants $A$ and $B$, we have studied behaviour of the effective energy density, pressure and effective EoS parameter where $A=B=1/2$. The dynamical properties of the reconstructed mimetic gravity are viable for a wide range of the constants $\alpha$ that correspond to the different models within nonmetricity function of choice. Our work closes with the discussion of the energy conditions associated with the analysed scenario. In this aspect, SEC is generally violated in the present ($z=0$) for a broad range of constants $\alpha$, a behaviour that aligns with the accelerated expansion that is currently observed.

In what follows, mimetic-$f(Q)$ (nonmetricity) gravity should be treated as an equivalent to the mimetic-$f(R)$ (curvature) and torsion mimetic-$f(T)$ (torsion) gravities \cite{nojiri2014m,mirza2019}. Thus, the geometrical trinity of mimetic gravity is completed and expanded to the arbitrary functional of nonmetricity, torsion or curvature \cite{beltran2019}. The overview and details of the new trinity will be done in the future works. Moreover, since the main scope of our work was associated with the introduction of the mimetic framework within STEGR along with the cosmological reconstruction and energy conditions, the many aspects of the theory introduced herein are yet to be studied. Specifically, in order to better address current cosmology, the models within mimetic-$f(Q)$ theory should be confronted with the current observations and $\Lambda CDM$, in order to constrain parameters of the theory and unveil additional insight \cite{lazkoz2019,xu2019f}.

Additionally, one can also extend the framework of mimetic-nonmetricity to the case where coupling between matter and nonmetricty is present, by considering $f(Q,T)$ extension of $f(Q)$ gravity \cite{gadbail2021,gadbail2023,narawade2023}. Then, the starting action will have the form:
\begin{align}
    S=\int \text{d}^4x\sqrt{-g}\frac{1}{2}\Big[f(Q,T)-U(\phi)+\eta(g^{\mu\nu}\partial_\mu\phi\partial_\nu\phi+1)  \Big]+\int\text{d}^4 x \sqrt{-g}\mathcal{L}_m,
\end{align}
from which the field equations can be derived similarly to the steps (\ref{eq8})-(\ref{eq13}). We will focus on the mimetic $f(Q,T)$ gravity in one of the future works.

In summary, the integration of mimetic theories based on nonmetricity introduces a novel and unexplored path toward further research avenues, offering additional degrees of freedom. The dark matter and dark energy are still mysterious, hence pursuing new models within modified gravity is still viable and theoretically motivated. These secrets may be unravelled in the new formulation of mimetic gravity. Moreover, since $f(Q)$ the theory is second order in metric, the STEGR framework may be more suitable in approaching various challenges normally present within higher-curvature formulations of (mimetic) gravity, such as emergence of ghosts \cite{sebastiani2017,hu2023}. Therefore, within the domain of mimetic-$f(Q)$ gravity, numerous implications—both theoretical and observational—remain uncharted, awaiting further investigations and studies.
\bibliographystyle{apsrev}
\bibliography{references}
\section{Appendix}
\subsection{Full form of $f(Q)$ function reconstructed from the Lagrange multiplier}
The integral of Eq.(\ref{eq43}) w.r.t. nonmetricity scalar $Q$ takes the form:
{\small\textrm{
  \setlength{\abovedisplayskip}{2pt}
  \setlength{\belowdisplayskip}{\abovedisplayskip}
  \setlength{\abovedisplayshortskip}{2pt}
  \setlength{\belowdisplayshortskip}{2pt}
\begin{align}
  &f(Q)=\frac{12}{t_0^3} \times\\ \nonumber &\Bigg[ \frac{2}{\Big(\sqrt{6} \sqrt{q} t_0-6 B\Big)^3 \Big(A \Big(-108 B^3 \gamma -54 \sqrt{6} B^2 \gamma  \sqrt{q} t_0-18 B t_0^2
   (\gamma  q-4 \eta_0)+\sqrt{6} \gamma  q^{3/2} t_0^3\Big)+12 c_1 \Big(6 B+\sqrt{6} \sqrt{q} t_0\Big)\Big)} \times \\ &
   \Big(A \Big(648 B^6 \gamma -108 B^4 t_0^2 (4 \eta_0+5 \gamma  q)+12 \sqrt{6} B^3 \sqrt{q} t_0^3 (18 \eta_0+7 \gamma 
   q)+18 B^2 q t_0^4 (\gamma  q-12 \eta_0)-6 \sqrt{6} B q^{3/2} t_0^5 (\gamma  q-2 \eta_0)\\ \nonumber &+\gamma  q^3
   t_0^6\Big)-12 c_1 \Big(6 B^2-q t_0^2\Big) \Big(6 B^2-2 \sqrt{6} B \sqrt{q} t_0+q t_0^2\Big)\Big) \Big(36
   \sqrt{6} A B \log \Big(\sqrt{6} B+\sqrt{q} t_0\Big) \Big(B^2 \gamma +\eta_0 t_0^2\Big)\\ \nonumber &+\sqrt{q} t_0 \Big(A \gamma
    \Big(-90 B^2-6 \sqrt{6} B \sqrt{q} t_0+q t_0^2\Big)+36 c_1\Big)\Big) \\ \nonumber &+\frac{\sqrt{\frac{2}{3}} A t_0 \Big(3 B^2 \gamma +\eta_0 t_0^2\Big) \Big(6 B \Big(\sqrt{6} B-\sqrt{Q} t_0\Big) \log
   \Big(\frac{A t_0 \Big(\sqrt{6} \sqrt{Q} t_0-6 B\Big)}{6 B^2-Q t_0^2}\Big)+\Big(6 \sqrt{6} B^2-12 B \sqrt{Q}
   t_0+\sqrt{6} Q t_0^2\Big) \tanh ^{-1}\Big(\frac{\sqrt{Q} t_0}{\sqrt{6} B}\Big)\Big)}{6 B^2-2 \sqrt{6} B \sqrt{Q}
   t_0+Q t_0^2} \\ \nonumber
   &-\frac{1}{6} A t_0 \Big(3 B^2 \gamma +\eta_0 t_0^2\Big) \times \\ \nonumber &\Big\{\frac{\Big(6 B-\sqrt{6} \sqrt{Q}
   t_0\Big) \Big(\Big(6 B-\sqrt{6} \sqrt{Q} t_0\Big) \log \Big(\sqrt{6} \sqrt{Q} t_0-6 B\Big)+6
   B\Big) \Big(\log \Big(\frac{A t_0 \Big(\sqrt{6} \sqrt{Q} t_0-6 B\Big)}{6 B^2-Q t_0^2}\Big)+\log
   \Big(\frac{\sqrt{6} B}{t_0}+\sqrt{Q}\Big)\Big)}{6 B^2-2 \sqrt{6} B \sqrt{Q} t_0+Q t_0^2}\\ \nonumber &-6
   \Big(\text{Li}_2\Big(\frac{1}{12} \Big(\frac{\sqrt{6} \sqrt{Q} t_0}{B}+6\Big)\Big)+\log \Big(\frac{\sqrt{6}
   B}{t_0}+\sqrt{Q}\Big) \log \Big(\frac{1}{2}-\frac{\sqrt{Q} t_0}{2 \sqrt{6} B}\Big)\Big)-\frac{6
   \sqrt{6} B \log \Big(\frac{\sqrt{6} B}{t_0}+\sqrt{Q}\Big)}{\sqrt{6} B-\sqrt{Q} t_0}+6 \tanh
   ^{-1}\Big(\frac{\sqrt{Q} t_0}{\sqrt{6} B}\Big)\Big\} \\ \nonumber &-\frac{1}{6} A t_0 \Big(3 B^2 \gamma +\eta_0 t_0^2\Big) \times \\ \nonumber &\Big\{-6 \Big(\text{Li}_2\Big(\frac{\sqrt{Q} t_0}{\sqrt{6} B}+1\Big)+\log \Big(\frac{\sqrt{6} B}{t_0}+\sqrt{Q}\Big) \log \Big(-\frac{\sqrt{Q} t_0}{\sqrt{6} B}\Big)\Big)+6 \Big(\text{Li}_2\Big(\frac{1}{12} \Big(\frac{\sqrt{6} \sqrt{Q}
   t_0}{B}+6\Big)\Big) \\ \nonumber &+\log \Big(\frac{\sqrt{6} B}{t_0}+\sqrt{Q}\Big) \log \Big(\frac{1}{2}-\frac{\sqrt{Q} t_0}{2 \sqrt{6} B}\Big)\Big)-\frac{6 \sqrt{6} B \log \Big(\frac{\sqrt{6} B}{t_0}+\sqrt{Q}\Big)}{\sqrt{6} B-\sqrt{Q} t_0}+6 \tanh
   ^{-1}\Big(\frac{\sqrt{Q} t_0}{\sqrt{6} B}\Big) 
   \\  \nonumber
 &+\frac{1}{6 B^2-2 \sqrt{6} B \sqrt{Q} t_0+Q t_0^2}\Big[  \Big(6 B-\sqrt{6} \sqrt{Q} t_0\Big) \Big(\Big(6 B-\sqrt{6} \sqrt{Q} t_0\Big) \log \Big(\sqrt{6} \sqrt{Q} t_0\Big) \\  &+\Big(\sqrt{6} \sqrt{Q} t_0-6 B\Big) \log \Big(\sqrt{6} \sqrt{Q} t_0-6 B\Big)+6 B\Big) \Big(\log \Big(\frac{A t_0
   \Big(\sqrt{6} \sqrt{Q} t_0-6 B\Big)}{6 B^2-Q t_0^2}\Big)+\log \Big(\frac{\sqrt{6} B}{t_0}+\sqrt{Q}\Big)\Big) \Big]   
   \Big\}
   \Bigg],
\end{align}}
where $Li_2(z)$ is Spence's (dilogarithm) function \cite{kirillov1995}.

\subsection{General behaviour of energy conditions}

\begin{figure}
    \centering
    \includegraphics[scale=0.92]{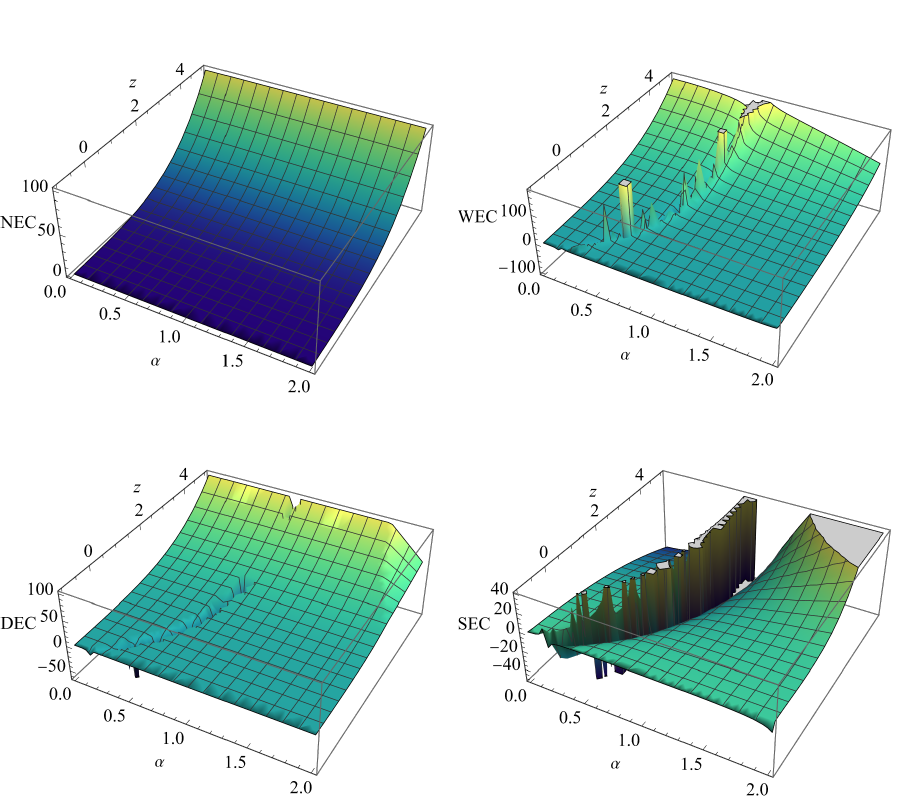}
    \caption{General behaviour of energy conditions across different values of redshift $z$ and $\alpha$}
    \label{figApp}
\end{figure}

\end{document}